\documentclass[twocolumn, a4paper, 10pt, showpacs, prx, reprint, aps, superscriptaddress, amsmath, amssymb, amsfonts, floatfix]{revtex4-1}
\usepackage{dcolumn}
\usepackage{bm}
\usepackage{graphicx, array}
\usepackage{rotating}
\usepackage[caption=false]{subfig}
\usepackage{xcolor}
\usepackage{float}
\usepackage{natbib}

\usepackage{changes}
\makeatletter
\newcommand\resetchangescolor[1]{%
  \setkeys{Changes@definechangesauthor}{color=#1}%
  \expandafter%
  \let\csname Changes@AuthorColor\endcsname=\Changes@definechangesauthor@color%
  \colorlet{Changes@Color}{\@nameuse{Changes@AuthorColor}}%
}
\makeatother

\resetchangescolor{red}

\begin{document}

\preprint{APS/123-QED}


\title{Two-dimensional transition metal chalcogenides with hexagonal and orthorhombic structures: candidates for auxetics and photocatalysts}

\author{Wenqi Xiong}
\affiliation{Key Laboratory of Artificial Micro- and Nano-structures of Ministry of Education and School of Physics and Technology, Wuhan University, Wuhan 430072, China}
\author{Kaixiang Huang}
\affiliation{Key Laboratory of Artificial Micro- and Nano-structures of Ministry of Education and School of Physics and Technology, Wuhan University, Wuhan 430072, China}
\author{Shengjun Yuan}
\email{s.yuan@whu.edu.cn}
\affiliation{Key Laboratory of Artificial Micro- and Nano-structures of Ministry of Education and School of Physics and Technology, Wuhan University, Wuhan 430072, China}

\begin{abstract}
In this paper, we perform theoretical study on the physical properties of two-dimensional transition metal chalcogenides MX$_{2}$ and M$_{2}$X$_{3}$ (M= Ni, Pd; X= S, Se, Te). These studied materials are classified in three stable phases according to their lattice structures: hexagonal MX$_{2}$, orthorhombic MX$_{2}$ and orthorhombic M$_{2}$X$_{3}$. They have either isotropic or anisotropic in-plane properties depending on their symmetries. In particular, the orthorhombic MX$_{2}$ and M$_{2}$X$_{3}$ have low lattice symmetry and present highly anisotropic properties. The orthorhombic MX$_{2}$ possess giant negative in-plane Poisson's ratios, different from the other two phases. Moreover, by joint analysis of band gap, band edge and optical absorption, the orthorhombic MX$_{2}$ and M$_{2}$X$_{3}$ are found to be highly efficient as water splitting photocatalysts within the visible and ultraviolet sunlight regions.
\end{abstract}


\maketitle


\section{INTRODUCTION}

Two-dimensional (2D) group-VIII transition metal chalcogenides (TMCs), including the transition metal dichalcogenides (for example, MoS$_{2}$, MoSe$_{2}$, WS$_{2}$ and WSe$_{2}$, etc.) have presented great potentials in electronic and optical devices, such as high current on/off ratio field-effect transistors (FETs) \cite{A3,A4}, photodetectors \cite{A5,A6} and valleytronic applications \cite{A7,A8}. Most reported 2D TMCs have isotropic mechanical and electronic properties due to their highly symmetric structures. The symmetry of lattice structure indeed plays a vital role in determining the electronic properties of materials. By lowering the symmetry of the structure, it is possible to induce strong in-plane anisotropic properties in 2D materials, as observed in puckered phosphorene \cite{A7+,A7++} and group-IV monochalcogenides (SnS, SnSe, GeS, GeSe) \cite{A9,A10,A11}. As have already been reported, the anisotropic properties have the advantage for certain applications, such as polarized light detection devices and valleytronics \cite{A11+,A11++,A11+++}.

Recently, few-layer PdSe$_{2}$ has been successfully synthesized via mechanical exfoliation and selenization on the precursor Pd layer \cite{A12,A13,A14}, which has aroused great interests due to its ambient stability, high carrier mobility ($\sim$158 cm$^{2}$V$^{-1}$s$^{-1}$) and in-plane anisotropic properties \cite{A15}. In contrast to its hexagonal phase (Fig.\ref{FIG1}(b)), the synthesized few-layer PdSe$_{2}$ forms an orthorhombic lattice with puckered pentagonal structure as illustrated in Fig. \ref{FIG1}(c). Moreover, Li $et~al.$ revealed a much lower diffusion barrier of Se vacancies in PdSe$_{2}$ than that of S vacancy in MoS$_{2}$ \cite{A16}. Further, Lin $et~al$. demonstrated that the introduction of Se vacancy in few-layer PdSe$_{2}$ can enhance the interlayer interaction and decrease the Se/Pd element ratio, which creates a new structure phase, $i.e.$, Pd$_{2}$Se$_{3}$ \cite{A17}, as shown in Fig. \ref{FIG1}(d). It is therefore highly desired to look for other 2D TMCs with a similar structure as PdSe$_{2}$ or Pd$_{2}$Se$_{3}$, which may be stable and reveal the in-plane anisotropic properties.

Furthermore, one important application of 2D materials is the photocatalytic water splitting, $i.e.$, to convert solar energy into chemical energy without additional cost \cite{A19,A20,A21}. Since the discovery of TiO$_{2}$ as a photocatalyst  in 1972 \cite{A18}, more and more semiconductors have been proposed, including 2D semiconductors such as g-C$_{3}$N$_{4}$ and MoS$_{2}$ \cite{18+,18++}. Particularly, one or few-layers 2D semiconductors are indeed thin, with inherent merit as photocatalysts due to their extremely large specific surface area. In fact, an ideal photocatalyst should have two characters: (i) the band edge need to straddle both the reduction potential of H$_{2}$/H$^{+}$  (-4.44 eV at pH=0) and the oxidation potential of H$_{2}$O/O$_{2}$ (-5.67 eV at pH=0); (ii) an appropriate optical gap to guarantee the absorption of the solar light. Once satisfying the principle (i), the holes and electrons can drive both the oxidation and reduction reactions to generate O$_{2}$ and H$_{2}$ from aqueous solution. We will use these principles to look for possible photocatalysts from the 2D TMCs studied in this paper.

In this paper, we will perform systematic study on 2D TMCs based on the VIII-VIA compounds in the form of M$_{m}$X$_{n}$, where M refers to the elements Ni and Pd,  and X represents S, Se and Te. We will calculate the physical properties of eighteen 2D TMCs from first principles by using density functional theory (DFT). The structures of these materials are classified in three phases, namely, the hexagonal MX$_{2}$ (H-MX$_{2}$), the orthorhombic MX$_{2}$ (O-MX$_{2}$) and the orthorhombic M$_{2}$X$_{3}$ (O-M$_{2}$X$_{3}$). In the following, We will first present the details of the numerical methods in Sec. II and show the main results in Sec. III, including stability, mechanical, electronic and optical properties of monolayer M$_{m}$X$_{n}$, and their potential applications in photocatalyst. Finally, we summarize our major findings In Sec. IV.

\section{CALCULATION METHOD}

The electronic properties of TMCs are calculated from first-principles by using DFT as implemented in VASP code \cite{A22}. The Perdew-Burke-Ernzerhof (PBE) parametrized generalized gradient approximation (GGA) and projected augmented wave (PAW) are adopted to describe exchange correlation potential and ion-electron interaction\cite{A23,A24}. The kinetic energy cutoff and $k$-point mesh of Brillouin zone (BZ) are set to 500 eV and 15$\times$15$\times$1 \cite{A25}, respectively. A vacuum thickness of 20 \AA{} is added to avoid the periodic interaction. Moreover, the energy convergence criteria and stress forces are set to 10$^{-5}$ eV and 0.01 eV/\AA{}, respectively. For few-layer TMCs, the van der Waals (vdW) force is corrected by using a semi-empirical DFT-D2 method \cite{A26,A27}. The spin-orbit coupling (SOC) is added into self-consistent calculations. Also, the corrected band structures are calculated by adopting hybrid Heyd-Scuseria-Ernzerhof (HSE06) method \cite{A28}.

The thermal stability of monolayer TMCs is evaluated by using PHONOPY code based on density functional perturbation theory (DFPT) and finite difference method \cite{A29,A30,A31}. We construct a 3$\times$3 supercell and adopt 5$\times$5$\times$1 $k$-point mesh to obtain force constants and phonon spectrum. In order to eliminate the imaginary frequency, the highly accurate energy convergence criteria and stress forces are set to 10$^{-8}$ eV and 10$^{-4}$ eV/\AA{}, respectively.

\section{RESULTS AND DISCUSSIONS}
\subsection{Structural stability}

\begin{figure}[]
	\centering
	\includegraphics[width=8.5cm]{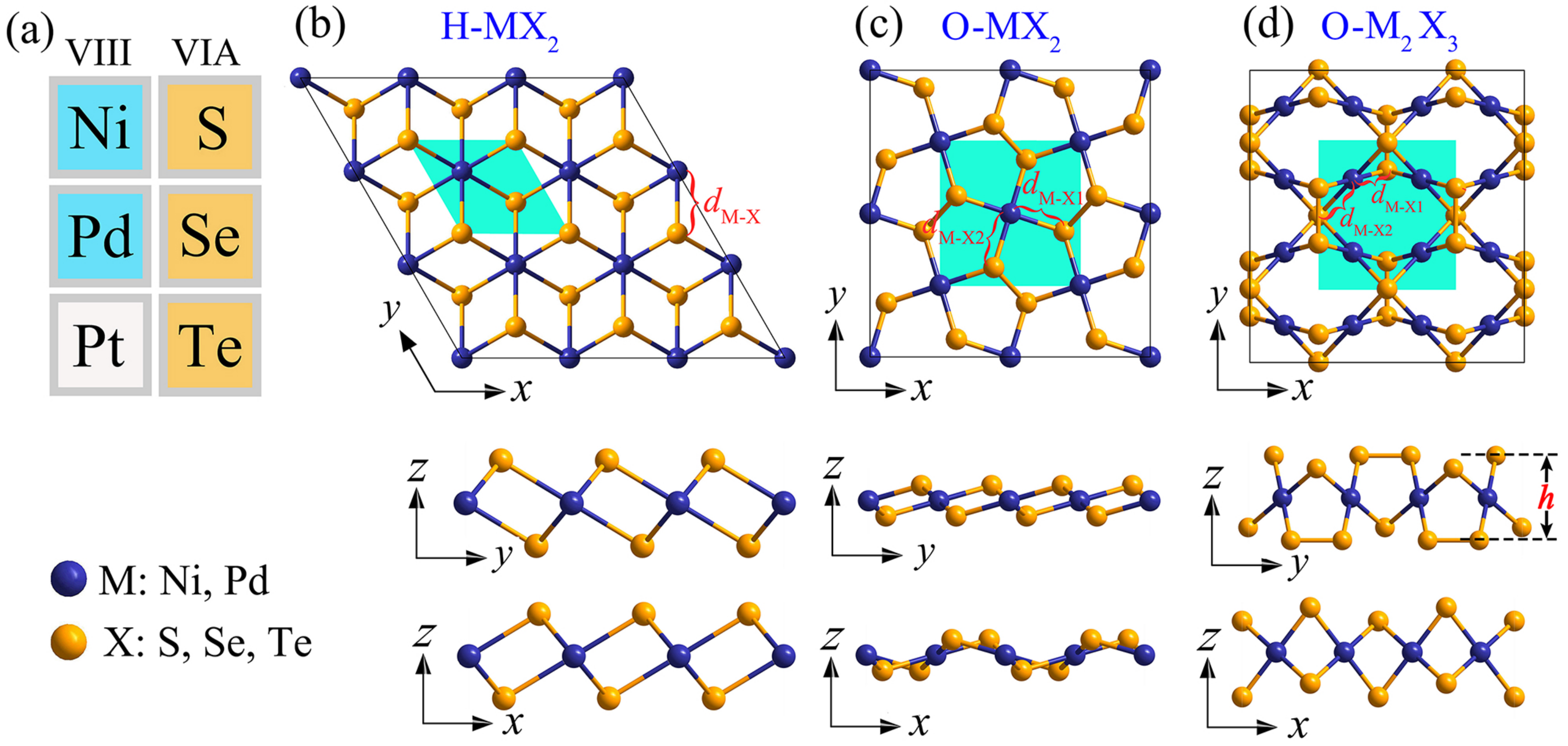}
	\caption{(a) The VIII and VIA elements in periodic table. Top and side views of (b) H-MX$_{2}$, (c) O-MX$_{2}$ and (d) O-M$_{2}$X$_{3}$, respectively. The green regions denote the unit cell.}
	\label{FIG1}
\end{figure}

We firstly study the basic geometric structures and thermal stability of monolayer M$_{m}$X$_{n}$ (M= Ni, Pd; X= S, Se, Te). After fully optimizing the atomic positions, the M$_{m}$X$_{n}$ stabilizes into three structural symmetries as shown in Fig. \ref{FIG1}, namely, the hexagonal H-MX$_{2}$, the orthorhombic O-MX$_{2}$, and the orthorhombic O-M$_{2}$X$_{3}$.  All geometric structures are built up with three atomic layers as X-M-X, in which one M layer is sandwiched with two X layers. Each M atom binds six X atoms in H-MX$_{2}$ and four X atoms in O-MX$_{2}$ and O-M$_{2}$X$_{3}$. Detailed geometry analysis show that the space groups of H-MX$_{2}$, O-MX$_{2}$ and O-M$_{2}$X$_{3}$ are $P$3$m$1 (No. 164), $P$2$_{1}$/$c$ (No. 14) and $Pmmn$ (No. 59), respectively. In contrast to the common phase of hexagonal H-MX$_{2}$, the space groups of O-MX$_{2}$ and O-M$_{2}$X$_{3}$ have much lower symmetry.

The relaxed structural parameters, such as lattice constants ($a$ and $b$), bond lengths ($d_{M-X}$), and vertical heights ($h$) are listed in Table 1. These results show clearly that when the atomic radius of element M (X) increases from Ni (S) to Pd (Te), all bond lengths within the same structural phase always increase. Comparing to the hexagonal structure of H-MX$_{2}$, the bond lengths in two orthorhombic O-MX$_{2}$ and O-M$_{2}$X$_{3}$ are more close to each other, but the height, defined as the out of plane distance between the top and bottom X sub-layers, are totally different. Furthermore, for O-MX$_{2}$ and O-M$_{2}$X$_{3}$ phases, there are diversity of in-plane lattice constants along different crystal lines, originating from the bond length difference between $d_{M-X1}$ and $d_{M-X2}$, as illustrated in Fig. \ref{FIG1}. These differences break further the geometry symmetry and induce subsequently anisotropic mechanical and optical properties, as we will explore in detail in the following.

\begin{table}
\renewcommand\arraystretch{1.2}
\centering
  \caption{Calculated lattice constants ($a$ and $b$), bond lengths ($d_{M-X}$), vertical heights ($h$) band gaps from PBE ($E_{gap}^{PBE}$) and HSE06 ($E_{gap}^{HSE}$) methods of monolayer TMCs.}
  \begin{tabular}{p{1.42cm}<{\centering}|p{1.0cm}<{\centering}p{1.0cm}<{\centering}p{1.5cm}<{\centering}p{1.2cm}<{\centering}p{0.8cm}<{\centering}p{0.8cm}<{\centering}}
  \hline\hline
      & $a$ (\AA{})&  $b$ (\AA{}) & $d_{M-X}$ ({\AA})  & $h$&$E_{gap}^{PBE}$ (eV) & $E_{gap}^{HSE}$ (eV)   \\
    \hline
    H-NiS$_{2}$ & 3.348	&3.348	&2.258	&  2.330 &0.61&	1.10\\
    H-NiSe$_{2}$ & 3.547&3.547&	2.390& 2.467 &0.21&	0.58 \\
    H-NiTe$_{2}$ & 3.787&3.787	&2.576&	2.721  &0&0 \\
    H-PdS$_{2}$ & 3.548	&3.548&	2.395 & 2.480  &1.27&1.80\\
    H-PdSe$_{2}$ & 3.730&3.730&	2.523& 2.627  &0.72	&1.13 \\
    H-PdTe$_{2}$ & 4.026&4.026&	2.701& 2.756 &0.26	&0.52 \\
    \hline
    O-NiS$_{2}$ & 5.215	&5.326&	2.172/2.182	& 1.149 &	0.82&2.40 \\
    O-NiSe$_{2}$ & 5.512&5.702&2.305/2.314& 1.368 &1.02&2.27\\
    O-NiTe$_{2}$ & 5.955&6.261&	2.489/2.498	& 1.552  &	0.95&1.89 \\
    O-PdS$_{2}$ & 5.472	&5.571&	2.328/2.339	&  1.267  &	1.18&2.14 \\
    O-PdSe$_{2}$ & 5.744&5.919&	2.452/2.462	& 1.488  &	1.36&2.16\\
    O-PdTe$_{2}$ & 6.146&6.439&	2.625/2.631& 1.693  &1.27&1.90 \\
\hline
    O-Ni$_{2}$S$_{3}$ & 5.239&5.57&2.190/2.256& 3.438  &0.38&	1.77 \\
    O-Ni$_{2}$Se$_{3}$ & 5.423&5.926&	2.310/2.383	& 3.704  &	0.37&	1.61 \\
    O-Ni$_{2}$Te$_{3}$ & 5.499&6.737&2.488/2.571& 4.105  & 0.30&	1.05 \\
    O-Pd$_{2}$S$_{3}$ & 5.773&5.907&2.341/2.427& 3.582  &0.45	&1.50 \\
    O-Pd$_{2}$Se$_{3}$ & 5.976&6.114&	2.455/2.539	& 3.842  &	0.42	&1.39 \\
    O-Pd$_{2}$Te$_{3}$ & 6.122&6.608&	2.622/2.696	& 4.228	 & 0.60&	1.24 \\
    \hline\hline
  \end{tabular}
\end{table}

\begin{figure*}[]
\centering
\includegraphics[width=13cm]{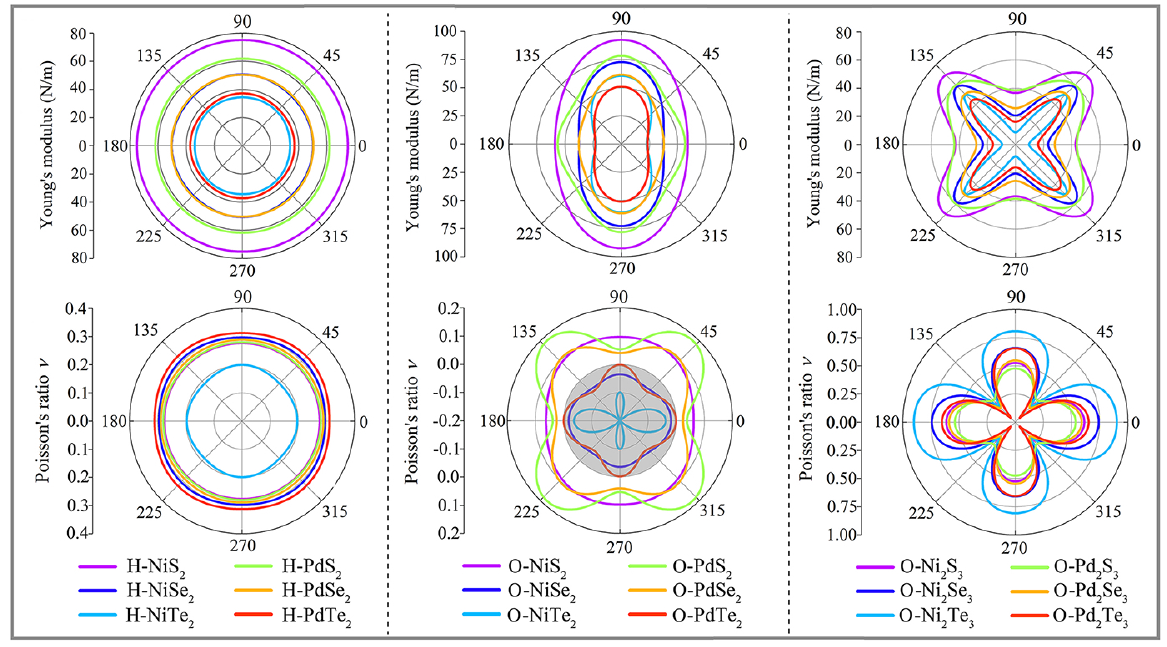}
\caption{Calculated orientation-dependent Young's modulus $Y(\theta)$ (top panel) and Poisson's ratio $v(\theta)$ (bottom panel) for H-MX$_{2}$, O-MX$_{2}$ and O-M$_{2}$X$_{3}$, respectively. The grey region denotes negative Poisson's ratio.}
\label{FIG2}
\end{figure*}

The thermal stability of O-MX$_{2}$ and O-M$_{2}$X$_{3}$ is qualitatively examined by ab initio molecular dynamics simulations implemented in GULP \cite{A32,A33}. Here we skip the discussion of H-MX$_{2}$, as these common hexagonal structures have already been synthesized successfully \cite{A34,A35,A36}. In our stability analysis, the 3$\times$3 supercell of monolayer TMCs is used, the time interval and time step of the testing period are set to be 5$\times$10$^{-12}$ s and 1$\times$10$^{-15}$ s, respectively. Our results show that the total energy of each structure considered in the O-MX$_{2}$ and O-M$_{2}$X$_{3}$ phases is oscillating persistently around a fixed value during the whole simulation (see the data of 300K presented in Fig. S1 and S2 of Supplementary Information). The stability of these structures is  further conformed from the spectra of phonon dispersions shown in Fig. S3 and Fig. S4 of Supplementary Information. No negative acoustic branch is observed for all the structures considered in this paper. These results obtained from ab initio molecular dynamics simulations and phonon dispersions from first-principles indicate that the monolayer O-MX$_{2}$ and O-M$_{2}$X$_{3}$  with M= Ni, Pd and X= S, Se, Te are all stable at room temperature. We will then continue the study by examining their mechanical and electronic properties, together with the phase of H-MX$_{2}$.

\subsection{Mechanical properties}

Three phases of monolayer TMCs considered in this paper present totally different lattice structures.  Materials belonging to the same phase of TMCs may have similar properties originating from the characters of their space group, but these from different phases should have significant differences in their physical properties.  As one of the most important mechanical properties, we exam first Young's modulus $Y(\theta)$ and Poisson's ratio $v(\theta)$ in the following.

Based on Hooke's law, the relationship between stiffness constants and modulus is given by

\begin{equation}
\centering
\begin{bmatrix}\sigma_{xx}\\\sigma_{yy}\\\sigma_{xy}\end{bmatrix}=\begin{bmatrix}C_{11}&C_{12}&0\\C_{12}&C_{22}&0\\0&0&C_{66}\end{bmatrix}\begin{bmatrix}\varepsilon_{xx}\\\varepsilon_{yy}\\2\varepsilon_{xy}\end{bmatrix},
\end{equation}
where the in-plane stiffness tensor $C_{ij}$ ($i$,$j$=1,2,6) is equal to the second partial derivative of strain energy $E_{s}$, which is obtained by
\begin{equation}
\centering
E_s=\frac12C_{11}\varepsilon_{xx}^2+\frac12C_{22}\varepsilon_{yy}^2+C_{12}\varepsilon_{xx}\varepsilon_{yy}+2C_{66}\varepsilon_{xy}^2  ,
\end{equation}
where the tensile strain is defined as $\varepsilon$=($a$-$a_0$)/$a_0$, here $a$ and $a_{0}$ are strained and unstrained lattice constants, respectively. Young's modulus $Y$ and Poisson's ratio $v$ can be expressed as functions of the in-plane stiffness tensors as\cite{A37}
\begin{small}
\begin{equation}
	\centering
Y_x=\frac{C_{11}C_{22}-C_{12}^2}{C_{22}},
\qquad
Y_y=\frac{C_{11}C_{22}-C_{12}^2}{C_{11}},
\end{equation}
\end{small}
\begin{small}
\begin{equation}
	\centering
	v_x=\frac{C_{12}}{C_{22}},
	\qquad\qquad\qquad
	v_y=\frac{C_{12}}{C_{11}},
\end{equation}
\end{small}
In fact, the anisotropic mechanical feature can be further checked by calculating the orientation-dependent Young's modulus $Y$ and Poisson's ratio $v$, which can be expressed as \cite{A38}
\begin{small}
\begin{equation}
\centering
Y\left(\theta\right)=\frac{C_{11}C_{22}-C_{12}^2}{C_{11}s^4+C_{22}c^4+({\displaystyle\frac{(C_{11}C_{22}-C_{12}^2)}{C_{66}}}-2C_{12})s^2c^2},
\end{equation}
\end{small}
\begin{small}
\begin{equation}
	\centering
v\left(\theta\right)=\frac{C_{12}(s^4+c^4)-(C_{11}+C_{22}-\frac{(C_{11}C_{22}-C_{12}^2)}{C_{66}})s^2c^2}{C_{11}s^4+C_{22}c^4+({\displaystyle\frac{(C_{11}C_{22}-C_{12}^2)}{C_{66}}}-2C_{12})s^2c^2},
\end{equation}
\end{small}
where s=sin$\theta$ and c=cos$\theta$.

\begin{figure}[b]
	\centering
	\includegraphics[width=7cm]{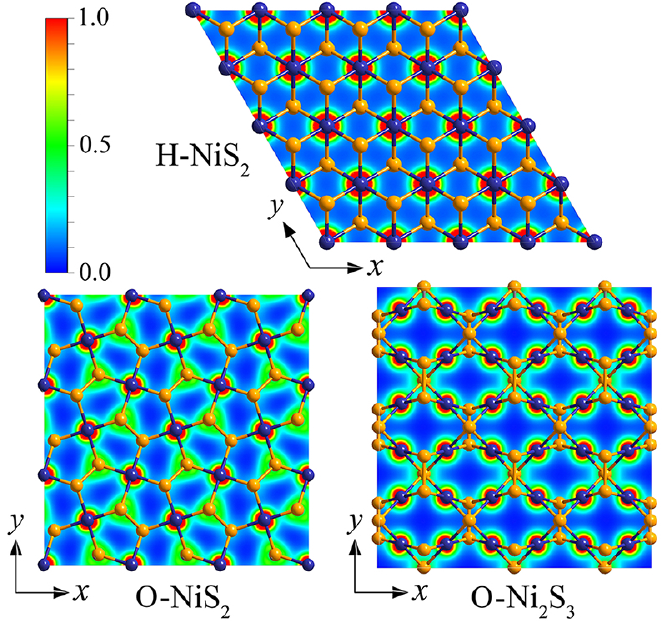}
	\caption{The distribution of the charge density in the ground states of H-NiS$_{2}$, O-NiS$_{2}$ and O-Ni$_{2}$S$_{3}$. The color indicates the relative amplitude of the local densities.}
	\label{FIG3}
\end{figure}

The in-plane stiffness tensors $C_{ij}$ are obtained from a series of strain $\mid$$\varepsilon$$\mid$ $\leqslant$ 2\% and a step of 0.5\%. All in-plane stiffness tensors $C_{ij}$ fitting from Eq. (2) for eighteen monolayer TMCs are collected in Table S1 of Supplementary Information. The orientation-dependent Young's modulus $Y(\theta)$ and Poisson's ratio $v(\theta)$ calculated by using Eq. (5) and (6) are plotted in Fig. \ref{FIG2}.  It is clear that all structures in the H-MX$_{2}$ phase are isotropic as both Young's modulus $Y(\theta)$ and Poisson's ratio $v(\theta)$ keep as constants when varying $\theta$; but the other two phases, O-MX$_{2}$ and O-M$_{2}$X$_{3}$, are highly anisotropic with clear angle-dependent mechanical properties.

Particularly, Young's modulus of O-MX$_{2}$ phase increase monotonically from a minimum Young's modulus along $x$ direction ($\theta$=0$^{\circ}$) to a maximum value along $y$ direction ($\theta$=90$^{\circ}$). However, the maximum and minimum values of O-M$_{2}$X$_{3}$ phase are located at 45$^{\circ}$ and 0$^{\circ}$ (90$^{\circ}$), respectively. For the same element M ($i.e.$, Ni, Pd), Young's modulus decreases as X changes from S to Te due to the increment of the M-X bond strength. Furthermore, our calculations show that O-Ni$_{2}$Te$_{3}$ and O-Pd$_{2}$Te$_{3}$ have ultra-low Young's modulus ($<$ 20 N/m), which are even lower than monolayer graphene (340 N/m) and MoS$_{2}$ (125 N/m)\cite{A39,A40}, indicating their enormous potential in flexible devices.

For Poisson's ratio, besides the quite interesting anisotropic feather appeared in O-MX$_{2}$ and O-M$_{2}$X$_{3}$, our calculations show that three monolayer TMCs, O-NiSe$_{2}$, O-NiTe$_{2}$ and O-PdTe$_{2}$, present negative Poisson's ratios. The absolute value of negative Poisson's ratio obtained among these materials is -0.228 in O-NiTe$_{2}$ along 56$^{\circ}$ to the $x$ axis (see Fig. \ref{FIG1}(c)). A material with negative Poisson's ratio exhibits an interesting auxetic effect, $i.e.$, it expands along one direction if stretched along another direction. Auxetic materials are highly desirable for tissue engineering, bulletproof vests and many other medical applications. As a comparison to existing auxetic 2D materials such as borophene, penta-graphene, tinselenidene, etc., we collect and list their Poisson's ratios together with current values of O-NiSe$_{2}$, O-NiTe$_{2}$ and O-PdTe$_{2}$ in Table 2. O-NiTe$_{2}$ has lowest Poisson's ratio among three studied materials, which is comparable to other reported auxetic 2D materials.

\begin{figure*}[]
	\centering
	\includegraphics[width=12cm]{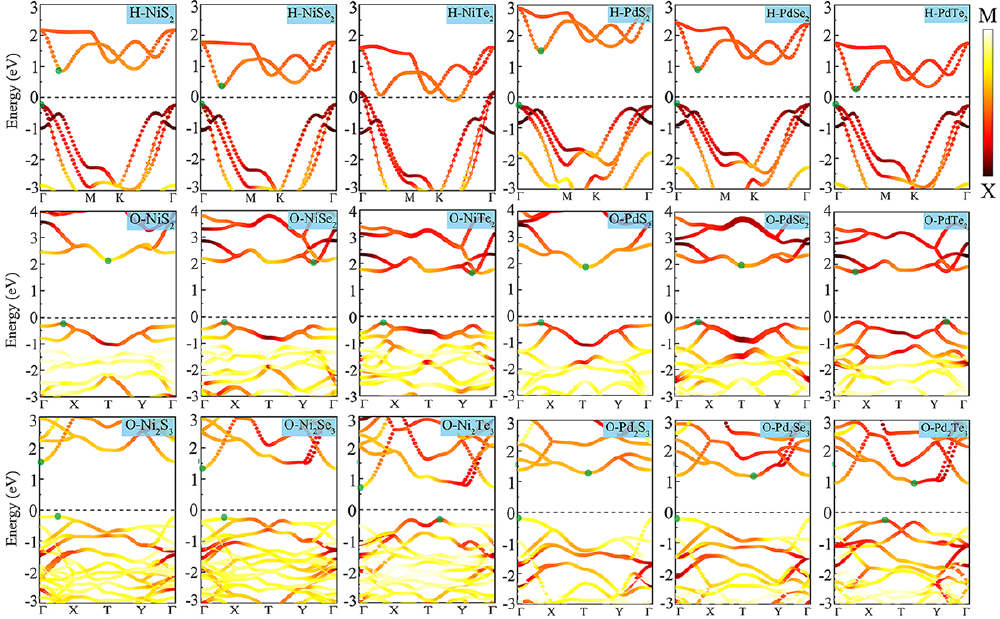}
	\caption{The HSE06 projected band structures of eighteen monolayer TMCs.}
	\label{FIG4}
\end{figure*}

Actually, the nature of isotropic or anisotropic mechanical properties can be explained by analyzing the charge densities obtained from first principles. Here we use the case of M=Ni and X=S as an example to compare the charge distributions in three different structural phases. As plotted in Fig. \ref{FIG3}, the charge densities of H-NiS$_{2}$ are localized isotropically around Ni atoms. On the contrary, the charge densities of O-NiS$_{2}$ and O-Ni$_{2}$S$_{3}$ are not uniformly distributed, but form patterns continuing along one crystal line. To be more precisely, for O-NiS$_{2}$, the extended pattern is along $y$ direction, and for O-Ni$_{2}$S$_{3}$, it is along the diagonal direction. This is, indeed, consistent with the calculated Young's modulus, in which the maximum value appear along the continuous pattern. The electron orbitals are hybridized stronger along these directions, leading to larger overlap of wave functions and larger bonding strength, and subsequently higher stiffness.

\begin{table}
\renewcommand\arraystretch{1.2}
\centering
  \caption{The negative Poisson's ratios $v$ in $x$, $y$ directions and its maximum values $v_{max}$ for other 2D materials.}
  \begin{tabular}{p{3cm}<{\centering}|p{1.4cm}<{\centering}p{1.4cm}<{\centering}p{1.4cm}<{\centering}}
  \hline\hline
      System  & $v_{x}$ & $v_{y}$ & $v_{max}$ \\
    \hline
    O-NiSe$_{2}$ & -0.018 & -0.036 & -0.050 \\
    \hline
    O-NiTe$_{2}$ & -0.037 & -0.100 & -0.228\\
    \hline\
    O-PdTe$_{2}$  & -0.001 & -0.002 & -0.058\\
    \hline
    Borophene \cite{A37} & -0.022 & -0.009 & - \\
    \hline
    $\delta$-Silica \cite{A41} & -0.123 & -0.112 & -\\
    \hline\
     Penta-graphene \cite{A42} & -0.068 & -0.068 & -\\
    \hline
     Be$_{5}$C$_{2}$ \cite{A43}& -0.041 & -0.16 & - \\
     \hline
      $\delta$-AsN \cite{A44} & -0.177& -0.068 & -0.296\\
    \hline
    Tinselenidene \cite{A45} & -0.171 & 0.46 & - \\
    \hline\hline
  \end{tabular}
\end{table}

\subsection{Electronic properties}

 \begin{figure}[]
 	\centering
 	\includegraphics[width=7cm]{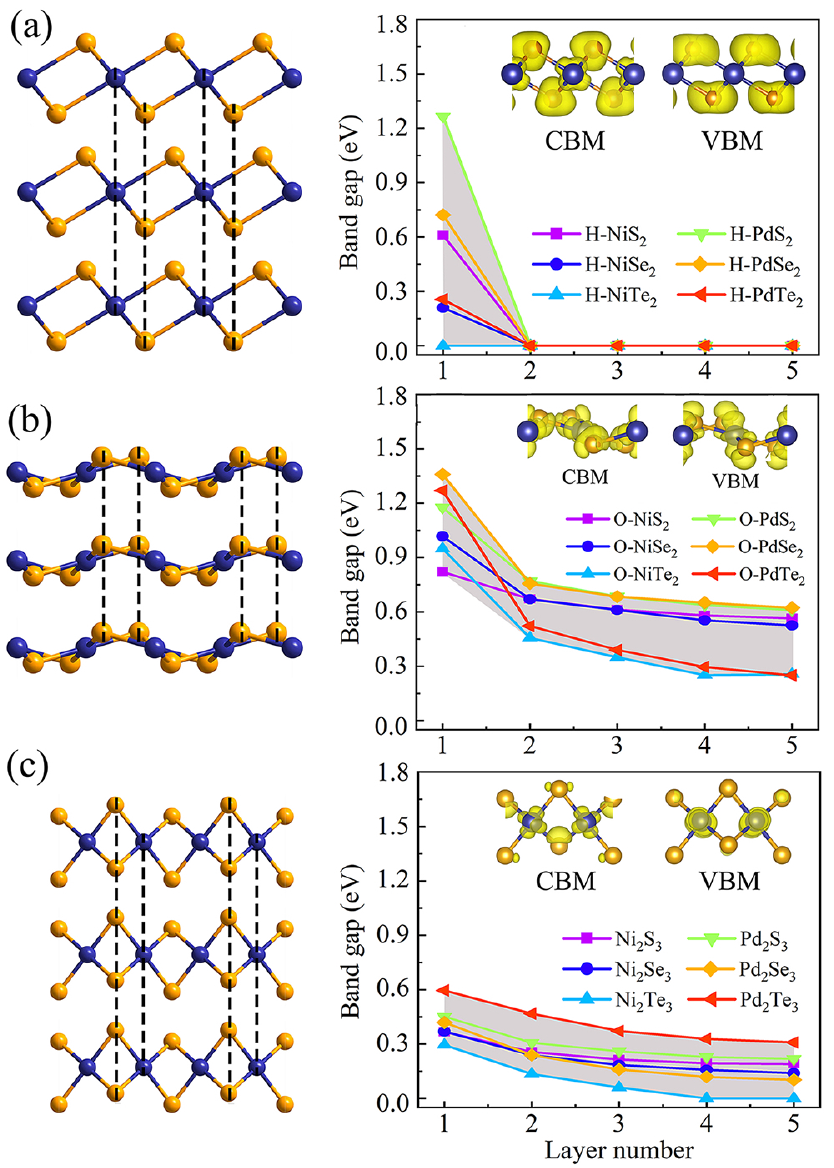}
 	\caption{Atomic structures of few-layer TMCs and layer number-dependent PBE band gaps for (a) H-MX$_{2}$, (b) O-MX$_{2}$ and (c) O-M$_{2}$X$_{3}$, respectively. The insets are band decomposed charge densities of monolayer H-NiS$_{2}$, O-NiS$_{2}$ and Ni$_{2}$S$_{2}$ for CBM and VBM, respectively.}
 	\label{FIG5}
\end{figure}

In this section, we study the electronic properties of monolayer TMCs. We firstly perform band structure calculations by using PBE. The results show that most materials considered in our paper are semiconductors (see details in Supplementary Information). As the PBE method usually underestimates the band gap of a semiconductor, we performed DFT calculations in VASP with more accurate HSE06 method and show the obtained band structures with projected densities in Fig. \ref{FIG4}. The HSE06 results are similar as those in PBE, and all monolayer TMCs are semiconductors with indirect band gaps,  except H-NiTe$_{2}$ which is a metal. Detailed analysis shows that, for H-MX$_{2}$ phase, the conduction band minimum (CBM) and the valence band maximum (VBM) are mainly attributed from element X; for O-MX$_{2}$ and O-M$_{2}$X$_{3}$ phases, the CBM and VBM originate from both compounds of M and X. The values of band gap obtained from both PBE and HSE06 are listed in Table 1. It indicates that for the materials in the same structural phase, the band gaps always decrease as the element X varies from S to Te, which are similar to those observed in MoX$_{2}$ and WX$_{2}$ (X=S, Se and Te) \cite{A45+}. As there are relatively heavy elements in the considered TMCs, it is worth checking also the effects of the SOC. From the results obtained with or without SOC in PBE (see details in Supplementary Information), the SOC interaction is overall negligible for most materials considered in our paper, except H-PdSe$_{2}$ and H-PdTe$_{2}$. In these two materials, there are large splittings of energy bands due to SOC, especially around the $\Gamma$ points. However, these splitting will not change qualitatively the properties we are interested in (see details in Fig. S7), therefore we will mainly show results without SOC in the following.

 All three phases of TMCs present stable multilayer structures stacked along the direction perpendicular to their plane. In Fig. \ref{FIG5}(a), we show atomic structures of stacked trilayer TMCs. The stacking sequence of H-MX$_{2}$, O-MX$_{2}$ and O-M$_{2}$X$_{3}$ is AAA stacking. We calculate further the electronic properties of multilayer TMCs by using relaxed structures as shown in Fig. \ref{FIG5}(a), and present the main results in Fig. \ref{FIG5}(b). Here, we consider mainly the thickness-dependence of the electronic properties of multilayer structure, and show the values of band gap with different number of layers ranging from 1 to 5.  As is well known, the interlayer vdW interaction, which is absent in a monolayer, plays a vital role to determine the properties of multilayer 2D materials, especially at low energy around the Fermi level. In general, the interlayer vdW interaction will lower the band gap for semiconducting 2D materials, because of the hybridization of the bands between neighboring layers. This is indeed also the case for TMCs considered in our paper. In particular, once H-MX$_{2}$ becomes a bilayer, its band gap promptly decreases to zero (see the results shown in Fig. S8), indicating that it gives a fierce response to the thickness. For O-MX$_{2}$ and O-M$_{2}$X$_{3}$ phases, their band gaps keep decreasing when adding more layers, but slowly. Specifically, the variation ranges of band gaps are 0 $\sim$ 1.27 eV, 0.26 $\sim$ 1.36 eV and 0 $\sim$ 0.60 eV for H-MX$_{2}$, O-MX$_{2}$ and O-M$_{2}$X$_{3}$, respectively. To further explore the origin of the relationship between the band gaps and the interlayer coupling, Fig. \ref{FIG5} shows the band decomposed charge densities of CBM and VBM in the monolayer. For H-NiS$_{2}$, O-NiS$_{2}$ and O-Ni$_{2}$S$_{3}$, charge densities of CBM and VBM are distributed among outside S atoms, Ni-S bonds and inside Ni atoms, respectively. When the monolayers are stacked together, the few-layer H-NiS$_{2}$ and O-Ni$_{2}$S$_{3}$ have maximum (minimum) interlayer charge overlapping, leading to maximum (minimum) change of band gap.

\subsection{Photocatalyst and light absorption}

Most monolayer TMCs considered in this paper are semiconducting with energy gaps ranging from 0.52 to 2.40 eV according to HSE06 calculations, providing a wide range of candidates for different optical applications. The main concern in the following is to study their potential applications in the photocatalytic water splitting, $i.e.$, converting the solar energy into the chemical energy without additional cost \cite{A18,A19,A20,A21}. As designing principles, a highly efficient water splitting photocatalyst should hold two characters: (i) a band gap about 2.0 eV for the harvest of the solar energy ; (ii) the band edges (CBM and VBM) must straddling both the reduction potential of H$_{2}$/H$^{+}$ (-4.44 eV, pH=0) and the oxidation potential of H$_{2}$O/O$_{2}$ (-5.67 eV, pH=0). Here the hydrogen production via photocatalytic water splitting needs ultrahigh solar energy harvest to drive the oxidation and reduction reactions.

To be more precise, in the oxidation reaction, the holes are used to generate O$_{2}$:
\begin{equation}
	\centering
	\mathrm{4h^++2H_2O\rightarrow O_2+4H^+},
\end{equation}
meanwhile, the excited electrons take part in the hydrogen reduction reaction to produce H$_{2}$:
\begin{equation}
	\centering
	\mathrm{4e^-+4H_2O\rightarrow 2H_2+4OH^-},
\end{equation}
In fact, the redox potentials of water are related to the pH of aqueous solution . According to the Nernst equation \cite{A46,A47,A48}, the water redox potentials and the value of pH satisfy the following relation:
\begin{equation}
	\centering
	E^\mathrm{pH}=E^{\mathrm{pH}=0}-0.059\times \mathrm{pH},
\end{equation}
which means that, the redox potentials of water increase linearly with pH by a factor of 0.059eV/pH.

\begin{figure}[htb]
	\centering
	\includegraphics[width=7cm]{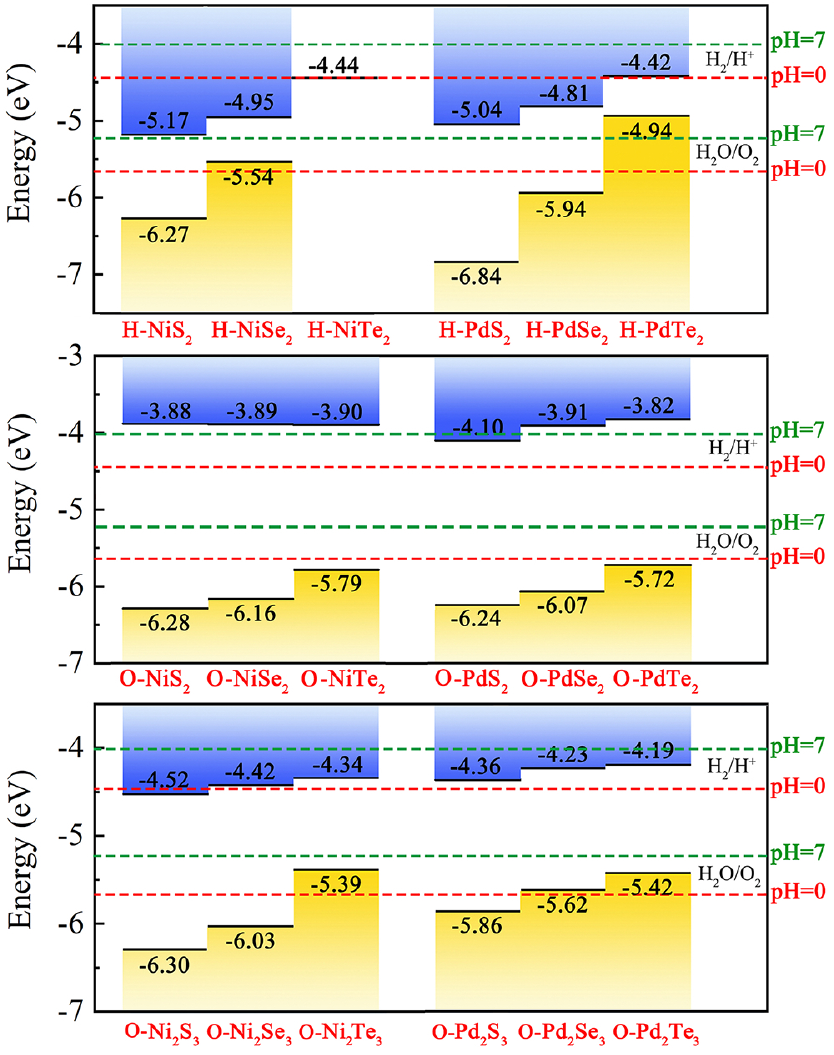}
	\caption{Band alignments of monolayer H-MX$_{2}$, O-MX$_{2}$ and O-M$_{2}$X$_{3}$ with respect to the redox potentials of water.}
	\label{FIG6}
\end{figure}

\begin{figure*}[htp]
	\centering
	\includegraphics[width=14cm]{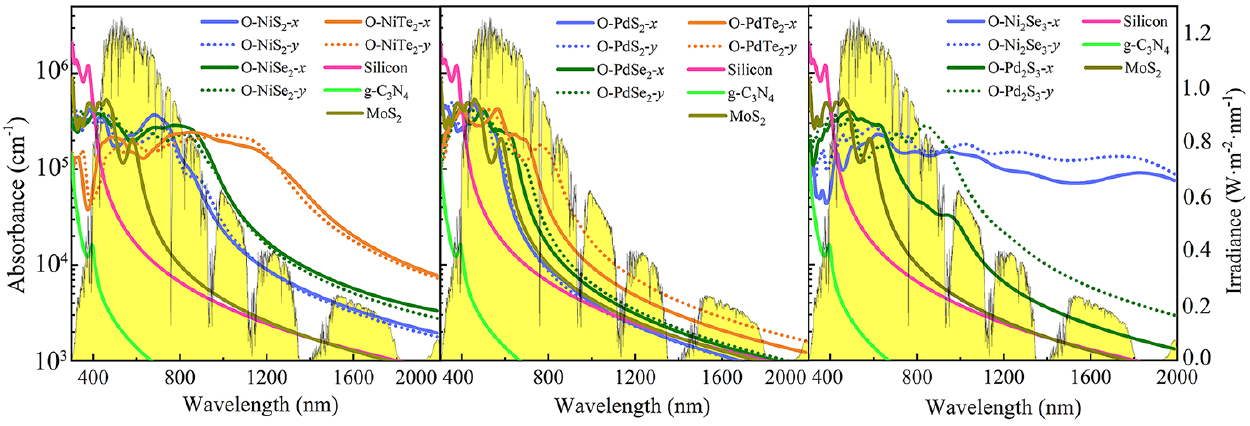}
	\caption{The absorption coefficients of monolayer O-MX$_{2}$ phase,  O-Ni$_{2}$Se$_{3}$ and O-Pd$_{2}$S$_{3}$, respectively, calculated with GW-BSE. The yellow background denotes the reference solar spectral irradiance in incident AM1.5G solar flux \cite{A49}.}
	\label{FIG7}
\end{figure*}

Here, in Fig. \ref{FIG6}, by adopting HSE06 method, the accurate band alignments of monolayer TMCs are obtained. For the H-MX$_{2}$ phase, the CBM and VBM never meet the requirement of redox potentials at pH=0 or 7, indicating that they can not be used for achieving water splitting. For the O-MX$_{2}$ phase, the CBM and VBM are always higher and lower than the reduction and oxidation potentials at pH=0, respectively, implying that they have inherent advantage in realizing water splitting. When the pH of aqueous solution  increases to 7, only the CBM of monolayer O-PdS$_{2}$ fails in producing H$_{2}$. For the O-M$_{2}$X$_{3}$ phase, both O-Ni$_{2}$Se$_{3}$ and O-Pd$_{2}$S$_{3}$ meet the redox potentials at pH=0, but they fail in realizing water splitting at pH=7. To further analysis the ability of water splitting, the kinetic overpotentials $\Delta$$E_{C}$ and $\Delta$$E_{V}$ (the difference between band edge and redox potential) are tested, which can represent properly the efficiency of driving the redox reaction (see details in Table S2 of the Supplementary Information). It shows that when pH of aqueous solution  increases, $\Delta$$E_{C}$ decreases and $\Delta$$E_{V}$ increases. The pH value-dependent kinetic overpotential shows tunable ability of H$_{2}$ production. These results imply that O-MX$_{2}$, O-Ni$_{2}$Se$_{3}$ and O-Pd$_{2}$S$_{3}$ are possible photocatalysts for water splitting at specific pH of aqueous solution .

To investigate the actual performance, we need further consider the sunlight harvest of these candidates by calculating optical absorption coefficients. Using the GW approximation in conjunction with the Bethe-Salpeter equation (BSE) \cite{A50,A51}, the  light absorbance is obtained and plotted in Fig. \ref{FIG7}. Here we include the electron-hole interaction in the optical calculation, as the charge screening effect is much weaker in two-dimension comparing to three-dimension due to the absence of screening along the out-of-plane direction. The solar energy is distributed in the infrared, visible and ultraviolet light about 43\%, 50\% and 7\%, respectively. Fig. \ref{FIG7} shows that O-MX$_{2}$,  O-Ni$_{2}$Se$_{3}$ and O-Pd$_{2}$S$_{3}$ have ultrahigh absorption coefficients within both visible (400-760 nm) and  ultraviolet ranges ($<$760 nm), indicating their excellent harvest of the solar energy. As a comparison, we perform optical calculations of widely used intrinsic silicon, and other 2D semiconductor photocatalysts, including g-C$_{3}$N$_{4}$ and MoS$_{2}$. When the wavelength is longer than 400 nm, the absorption coefficients of our TMCs candidates are much higher than all other compared materials, for example, about ten times higher than the value of intrinsic silicon. specially, O-Ni$_{2}$Se$_{3}$  shows high and constantly absorption over the entire energy range of the sunlight. Our results identify that O-MX$_{2}$, O-Ni$_{2}$Se$_{3}$ and O-Pd$_{2}$S$_{3}$ have large absorption coefficients from visible to ultraviolet light and provide congenital advantages for applications as photocatalyst. Furthermore, monolayer O-MX$_{2}$, O-Ni$_{2}$Se$_{3}$ and O-Pd$_{2}$S$_{3}$ present highly anisotropic optical properties, consistent with their mechanical properties. These materials can be used also for polarization-dependent photodetectors, similar as these proposed for other 2D materials such as black phosphorus \cite{A52}.

\section{CONCLUSIONS}
In conclusion, we have studied three phases of monolayer transition metal chalcogenides H-MX$_{2}$, O-MX$_{2}$ and O-M$_{2}$X$_{3}$ (M= Ni, Pd; X= S, Se, Te). We systematically examined their structural, mechanical and electronic characteristics via first-principle calculations. All these structures are stable at room temperature, verified by time-dependent ab initio molecular dynamics simulations and their phonon dispersion. The calculated mechanical properties also show that H-MX$_{2}$ is isotropic, while O-MX$_{2}$ and O-M$_{2}$X$_{3}$ present highly in-plane anisotropy due to their reduced lattice symmetry. Furthermore, O-MX$_{2}$ shows great auxeticity with giant negative in-plane Poisson's ratios, which are comparable to other known two-dimensional materials. Hence, O-MX$_{2}$ has ultra-low Young's modulus. By calculating the band alignments and light absorption coefficients, we concluded that O-MX$_{2}$, O-Ni$_{2}$Se$_{3}$ and O-Pd$_{2}$S$_{3}$ can be used as flexible water splitting photocatalysts within visible and ultraviolet light regions, because of their suitable band gaps, band edges and ultrahigh sunlight absorption.

\section*{ACKNOWLEDGEMENTS}
This work is supported by the National Key R$\&$D Program of China (Grant No. 2018FYA0305800). Numerical calculations presented in this paper have been performed on a supercomputing system in the Supercomputing Center of Wuhan University.

\bibliography{bibfile}

\end{document}